\documentstyle[preprint]{aastex}

\begin{document}

\title{Universal Behavior of Phase Correlations in Non-linear Gravitational Clustering}
\author{Peter Watts and Peter Coles}
\affil{School of Physics \& Astronomy, University of
Nottingham, University Park, Nottingham NG7 2RD, United Kingdom}
\email{peter.watts@nottingham.ac.uk, peter.coles@nottingham.ac.uk}
\author{Adrian Melott}
\affil{Department of Physics and Astronomy, University of Kansas, Lawrence, KS 66045}
\email{melott@kusmos.phsx.ukans.edu}
\newpage
\begin{abstract}
The large-scale structure of the Universe is thought to evolve by
a process of gravitational amplification from low-amplitude
Gaussian noise generated in the early Universe. The later,
non-linear stages of gravitation-induced clustering produce phase
correlations with well-defined statistical properties. In
particular, the distribution of phase differences $D$ between
neighboring Fourier modes provides useful insights into the
clustering phenomenon. Here we develop an approximate theory for
the probability distribution $D$ and test it using a large battery
of numerical simulations. We find a remarkable universal form for
the distribution which is well described by theoretical arguments.
\end{abstract}


\section{INTRODUCTION}

The large-scale structure of the Universe is a complex
interconnecting pattern whose structural elements comprise
filaments, sheets and clusters of galaxies surrounding large voids.
According to standard theories this ``cosmic web'' develops by a
process of gravitational instability from small initial
fluctuations in the density of a largely homogeneous early
Universe.

The physical description of an inhomogeneous Universe revolves
around the dimensionless density contrast, $\delta({\bf x})$,
which is obtained from the spatially-varying matter density
$\rho({\bf x})$ via $\delta ({\bf x}) = [\rho({\bf
x})-\rho_0] / \rho_0$ where $\rho_0$ is the global mean density. It
is useful to expand the density contrast in Fourier series, in
which $\delta$ is treated as a superposition of plane waves:
\begin{equation}
\delta ({\bf x}) =\sum \tilde{\delta}({\bf k}) \exp(i{\bf k}\cdot
{\bf x})\label{eq:FT}.
\end{equation}
The Fourier transform $\tilde{\delta}({\bf k})$ is complex and
therefore possesses both amplitude $C({\bf k})=|\tilde{\delta}
({\bf k})|$ and phase $\Phi_k$ where
\begin{equation}
\tilde{\delta}({\bf k})=C({\bf k})\exp(i\Phi_k)=A({\bf k}) + i
B({\bf k}); \label{eq:fourierex}
\end{equation}
the real and imaginary parts are $A$ and $B$ so that $ A({\bf k})
 =  C({\bf k})\cos(\Phi_k)$ and $B({\bf k}) = C({\bf
k})\sin(\Phi_k)$. We also have $C^2({\bf k})=A^2({\bf k})+B^2({\bf
k})$ and $\tan \Phi_k=B({\bf k})/A({\bf k})$.

In theories of structure formation involving cosmic inflation 
\citep{guth81,guthpi82,linde82,albstein82}, the initial fluctuations that seeded the
structure formation process form a Gaussian random field
\citep{bbks} possessing the properties of statistical homogeneity
and isotropy. In such fields the real and imaginary parts of
$\tilde{\delta}({\bf k})$ are independent Gaussians so that the
modulus $C({\bf k})$ has a Rayleigh distribution and the phases
$\Phi_k$ are uniformly random on the interval $[0,2\pi]$.
Obviously the distribution of $\delta$ in position space will also
be Gaussian; indeed all finite-dimensional joint probabilities of
$\delta$ in different locations are multivariate Gaussian in this
case \citep{bbks}.

Even if the primordial density fluctuations were indeed Gaussian,
the later stages of gravitational clustering must induce some form
of non-linearity. One particular way of looking at this issue is
to study the behavior of Fourier modes of the cosmological density
field. If the hypothesis of primordial Gaussianity is correct then
these modes began with random spatial phases. In the early stages
of evolution, the plane-wave components of the density evolve
independently like linear waves on the surface of deep water. As
the structures grow in mass, they interact with other in
non-linear ways, more like waves breaking in shallow water. These
mode-mode interactions lead to the generation of coupled phases.
While the Fourier phases of a Gaussian field contain no
information (they are random), non-linearity generates non-random
phases that contain much information about the spatial pattern of
the fluctuations but which is ignored entirely in the usual
clustering descriptors, such as the power spectrum (see below).
There have been a number of attempts to gain quantitative insight
into the behavior of phases in gravitational systems. Some studies
\citep{rydgram91,sodsut92,jainbert98} have concentrated on the evolution of phase shifts
for individual modes using perturbation theory and numerical
simulations. An alternative approach was adopted by 
\citet{scheretal91}, who developed a practical
method for measuring the phase coupling in random fields that
could be applied to real data. More recent studies have
established connections between phase evolution, clustering
dynamics and morphology \citep{chiangcoles00,chiang01,chiangetal02} and new methods have been
developed for visualizing phase information \citep{coleschiang00}.
Connections have also been demonstrated \citep{wattscoles03} between phase
information and alternative measures of non-linear clustering such
as the bispectrum \citep{scoccetal99,verdeetal00,verdeetal01}.

One of the major barriers to the more widespread use of
phase-based methods for probing cosmic nonlinearity is the lack of
any well-established statistical framework for quantifying the
information contained in the Fourier phases. Since phases are
angular variables traditional statistical measures of location and
dispersion are inappropriate. Moreover, probability distributions
for circular variables have very different properties to those for
variables defined on the real line; see the monographs by
\citet{mardia} and \citet{fisher} for more detailed
discussion. The upshot of this is that it is not yet known whether
there are useful standard distributions and limit theorems like
those that make the Gaussian distribution so useful and so
ubiquitous for statistical analysis. It is this deficiency we
address in this letter.

\section{STATISTICAL DESCRIPTION}

Non-linear clustering is difficult to handle rigorously with
analytic methods. Our approach is therefore to develop an
approximate theory and test it using numerical experiments. The
simulations we shall be comparing to theory are constructed within
a finite cubic volume with periodic boundary conditions. The
Fourier representation of the clustering pattern they reveal is
therefore discrete. In particular phases are defined for
wave-vectors on a cubic lattice which, for simplicity, we take to
have an integer spacing. In previous analyses of phase coupling
phenomena \citep{chiangcoles00,chiang01,chiangetal02,coleschiang00} it has been 
found useful to introduce a quantity $D_k$, defined by
\begin{equation}
D_k\equiv\Phi_{k+1}-\Phi_{k} \label{eq:D},
\end{equation} which measures the difference in phase of modes with
neighboring wave-numbers in one dimension. This has many
advantages as a  descriptor, not the least of which is that a
translation of the origin by $x$ which alters each $\Phi_k$ by
$kx$ only produces a constant offset in $D_k$. One can analyze a
three-dimensional simulation by extracting a vector $(D_{1},
D_{2}, D_{3})$ from differences in three orthogonal $k$-space
directions; the result contains information about both the
location and structure of features in $(x,y,z)$ space. One can
think of $D_k$ as a discrete representation of $d\Phi_k/d k$, the
phase gradient. Note that if the two angles $\Phi_1$ and $\Phi_2$
are independent and uniformly random then the difference
$\Phi_1-\Phi_2$ is also uniformly random, so that $D_k$ will be
uniformly distributed for Gaussian fields.

When fluctuations are small $(\delta<<1)$ they evolve linearly: the
initial statistics are preserved in this regime because each mode
evolves independently. When $\delta \sim 1$, however,
mode-coupling terms alter the distributions of both amplitudes and
phases. Strictly speaking, therefore, the real and imaginary parts
of the Fourier representation of $\delta$ in this regime are no
longer Gaussian. However the form of the Fourier expansion
(\ref{eq:FT}) itself guarantees that, as long as the
autocorrelations of $\delta$ do not have too large a spatial
extent, the Fourier superposition will be approximately Gaussian.
It is therefore still a reasonable approximation to take $A$ and
$B$ to be Gaussian even for non-linear systems \citep{fanbard95}. More
important for our purposes is the fact that, while for Gaussian
fields each $k$-mode is statistically independent,  a non-linear
field contains mode-mode correlations \citep{scocetal99a}. These induce
phase correlations which manifest themselves in departures from
uniformity of the distribution of $D_k$.

We can model the effect by assuming that the two neighboring modes
involved in (\ref{eq:D}) have real and imaginary parts that are
Gaussian distributed. For notational ease we suppress the label
${\bf k}$ and so the real parts of each are written $A_1$ and
$A_2$ and the imaginary parts $B_1$, $B_2$. We assume these can be
approximated as Gaussians but we allow them to be
cross-correlated. The four variables $(A_1, B_1, A_2, B_2)$
therefore possess a four-dimensional multivariate Gaussian
distribution.

If a set of random variables $X_1\ldots X_N$ have a multivariate
Gaussian distribution the joint probability density function of
the variables has the form
\begin{equation}
{\cal P}_N (X_1, ...,X_N) = \frac{1}{K} \exp \Bigl( -{1\over 2}
\sum_{i,j} X_i ~ M_{ij}^{-1} ~ X_j \Bigr),
\label{eq:MG}\end{equation} where $K=(2 \pi)^{N/2} ({\rm det}~
M)^{1/2}$ and the correlation matrix $M_{ij} = \langle X_i X_j
\rangle$. For the case we are interested in we can write $ \langle
A_i^2 \rangle=\langle B_i^2 \rangle = \sigma_i^2$ for $i=1,2$. The
quantity $\sigma^2(k)$ is related to the power-spectrum,
$P(k)$ defined by $P(k)=\langle |\tilde{\delta}({\bf
k})|^2\rangle$. We can also define a quantity $\alpha$ by $
\langle A_i A_j \rangle =\langle B_i B_j \rangle \equiv
\sigma_i\sigma_j \alpha$ where $\alpha({\bf k})$ parameterizes the
cross-correlation of the modes. We also have $\langle A_i B_i
\rangle = 0$ (no summation), guaranteeing translation invariance
\citep{chiangcoles00}, while $\langle A_i B_j \rangle =- \langle B_i
A_j\rangle \equiv \sigma_i\sigma_j \beta$ in a similar fashion to
$\alpha$. Note that $\alpha$ and $\beta$ may well be equal but we
have kept the most general possible form here. These two
parameters allow us to construct the required distribution for
${\cal P}(A_1, A_2, B_1, B_2)$. We now convert the distribution
$\cal P$ of the vector $(A_1, A_2, B_1, B_2)$ to the distribution
${\cal F}$ of the vector $(C_1,\Phi_1,C_2,\Phi_2)$. The result is
\begin{equation}
{\cal
F}=\frac{C_1C_2}{4\pi\sigma_1^2\sigma_2^2\left(1-\chi^2\right)}\exp(-Q/2),
\end{equation}
where \begin{equation} Q= \frac{\sigma_2^2C_1^2+\sigma_1^2C_2^2 -
2\sigma_1\sigma_2C_1C_2 \chi
\cos(\Phi_1-\Phi_2+\theta)}{\sigma_1^2\sigma_2^2
\left(1-\chi^2\right)}.
\end{equation}
In these expressions we have used $\chi^2=\alpha^2+\beta^2$ and
$\tan \theta=\alpha/\beta$. Note that if $\chi=0$ (no mode
correlations) this reduces to the product of two Rayleigh
distributions for $C_1$ and $C_2$ and two uniform distributions of
$\Phi_1$ and $\Phi_2$.

The next step is to construct a conditional distribution of one of the
phase angles, say $\Phi_2$, given specific values for the other three
variables, which we designate as $c_1, c_2$ and $\phi_1$.  This can be done
straightforwardly using Bayes' theorem. The result is
\begin{equation}
{\cal G}(\Phi_2|c_1,\phi_1, c_2) = \frac{1}{2\pi I_0(\kappa)} \exp \left[ - \kappa
\cos(\phi_1-\Phi_2 +\theta) \right] ,
\end{equation}
where $I_0$ is a modified Bessel function of order zero and in
which $\kappa = c_1c_2\chi/\sigma_1\sigma_2\left(1-\chi^2\right)$.
This conditional distribution shows how the correlation between
neighboring phases arises since it depends upon $\phi_1$. It is
deficient, however, in that this distribution also depends on
specified values for the amplitudes $c_1$ and $c_2$.

A better theory might be obtained by marginalizing over these
variables, but the integrals involved are messy. On the other hand the
distribution of amplitudes is relatively narrow, peaking around
$c_1=\sigma_1$ and $c_2=\sigma_2$ and $\chi$ is presumably small for
quasi-nonlinear stages. Under these circumstances we expect the
distribution of phase differences formed over a large number of pairs
to follow the form given above, i.e.
\begin{equation}
P(D)= \frac{1}{2\pi I_0(\kappa)} \exp \left[ - \kappa \cos(D-\mu)
\right] \label{eq:vm}.
\end{equation}
where $\mu$ is the mean angle.
This distribution is well-known in the field of circular
statistics where it is known as the {\em von Mises} distribution
\citep{mardia,fisher}. The mean, $\mu$, is controlled by the
positions of individual features of the distribution and will
consequently vary from sample to sample. The parameter $\kappa$,
related to $\chi$, describes the level of nonlinearity. When
$\kappa\rightarrow 0$ is small the distribution is approximately
uniform, while for small $\kappa$ it takes the form $P(D)\simeq
\frac{1}{2\pi} [ 1+ \kappa \cos (D-\mu) ]$ showing that initial
departures from random phases manifest themselves as a sinusoidal
perturbation of P(D). In the limit $\kappa\rightarrow\infty$ the
distribution tends towards a single spike at $\theta=\mu$; this
corresponds to a single concentration in position space.

\begin{figure}[c]
\epsscale{1.0}   
\plotone{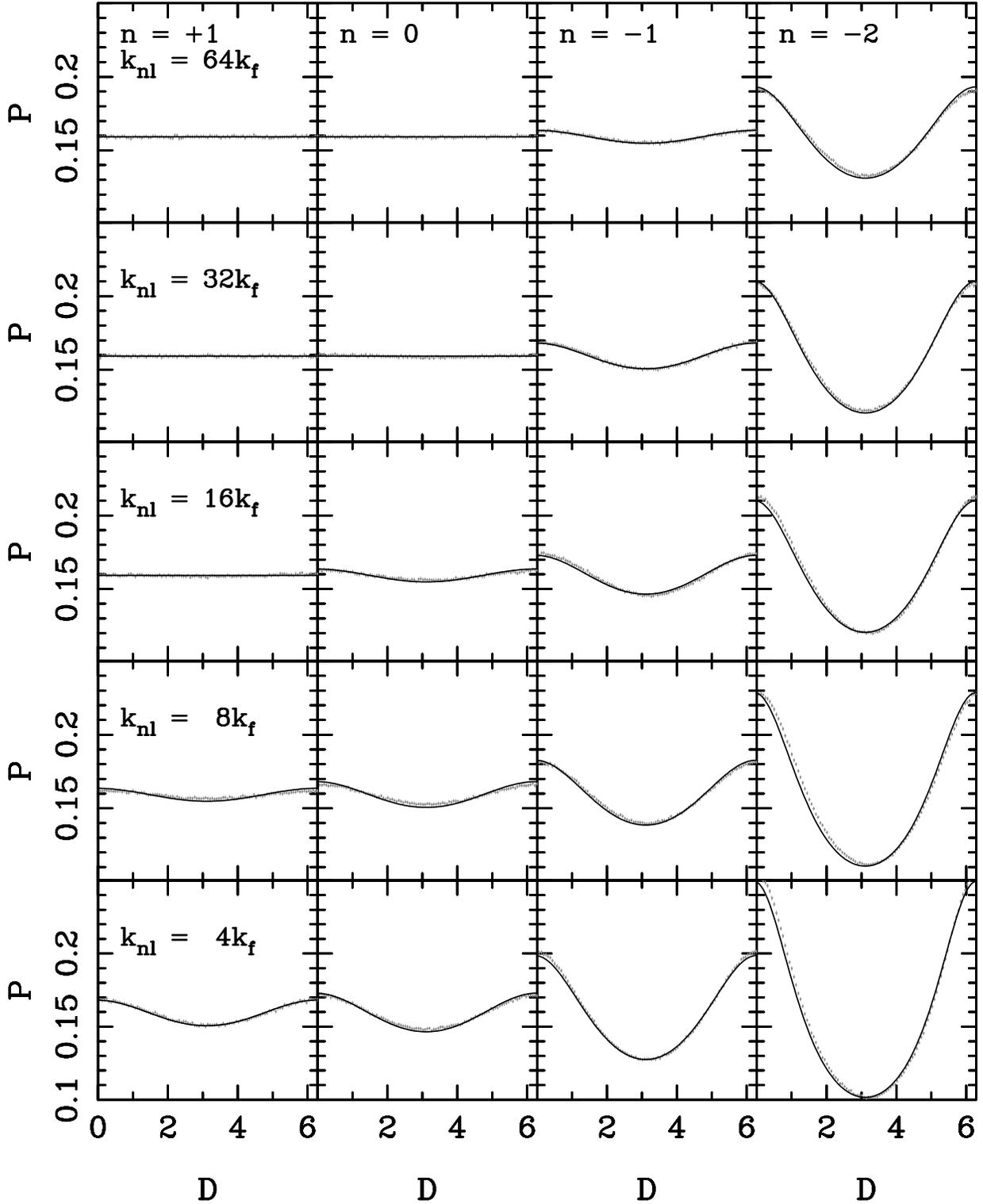}
\caption[]{Histograms $P(D)$ of phase differences $D$ are shown in grey  along with the
analytical model (\ref{eq:vm}) for four different initial spectra
as functions of epoch. Close agreement makes it difficult to discriminate between data and 
theory.\label{universe}}
\end{figure}

\placefigure{universe}

\section{NUMERICAL SIMULATIONS}

To test our hypothesis with fairly complete coverage of possible
parameters, we used a large ensemble of gravitational clustering
simulations \citep{melshand93}. These comprise sets of $128^3$ particles
and are not large by current standards, but the particular benefit
they offer our analysis is a fairly complete coverage of parameter
space. There are four realizations of each type of initial
conditions, using different pseudo-random number generators to
generate the initial phases. Initial power spectra were pure power
laws, i.e. $P(k) =A k^n$ with indices $n= -2$, $-1$, $0$, and $1$.
Data is taken every time the scale of clustering doubles, from
$k_{\rm nl} = 64 k_f$ to $k_{\rm nl} = 4 k_f$, where $k_f$ is the
fundamental mode of the box and $k_{\rm nl}$ is defined by
\begin{equation} \int_{0}^{k_{\rm nl}} P(k) d^3k=1. \end{equation}
The first stage may be significantly compromised by resolution
effects; we have simulations with $k_{\rm nl}=2k_f$ but these
almost certainly suffer from problems connected with the boundary
conditions (which are, as usual, periodic).

Analysis of the properties of the phase differences was conducted
on each realization of each spectrum and evolutionary phase, i.e.
a total of 120 times altogether. We Fourier-transform each stage
and extract phases for each wave-vector ${\bf k}$. From these we
obtain differences $D_k$ in three orthogonal $k$-space directions
from which we form histograms. The mean value $\mu$ of each
distribution contains information about the specific spatial
location of dominant features \citep{chiangcoles00}, which will differ from
realization to realization and which also varies with direction.
We therefore rotate individual distributions so that they have the
same mean value and ``stack'' the resulting histograms. We also
combine all three directional differences into an overall
histogram $P(D)$ where $D=(D_1^2+D_2^2+D_3^2)^{1/2}/\sqrt{3}$.
This approach, together with the large size of the ensemble,
produces final histograms with relatively small error bars, as can
be seen in Fig \ref{universe}.
Some readers may find it difficult to see the data, as it overlays
the theoretical model with high precision.

\placefigure{cdm}

\section{DISCUSSION}

The results show extremely good agreement over the range of initial
power-spectra and evolutionary stages, although the model does break
down at late stages for the case $n=-2$. This is not surprising, given
the fact that such spectra have large amounts of power on large scales
and phase correlations therefore develop extremely quickly. Even the
earliest stage shown of this simulation shows a significantly
non-uniform distribution of $D$.  The development of phase
correlations with evolutionary stage for each initial spectrum is
represented by the increasing deviation from uniformity down each
column starting, as expected, with sinusoidal departures. 

\begin{figure}[c]
\epsscale{0.8}   
\plotone{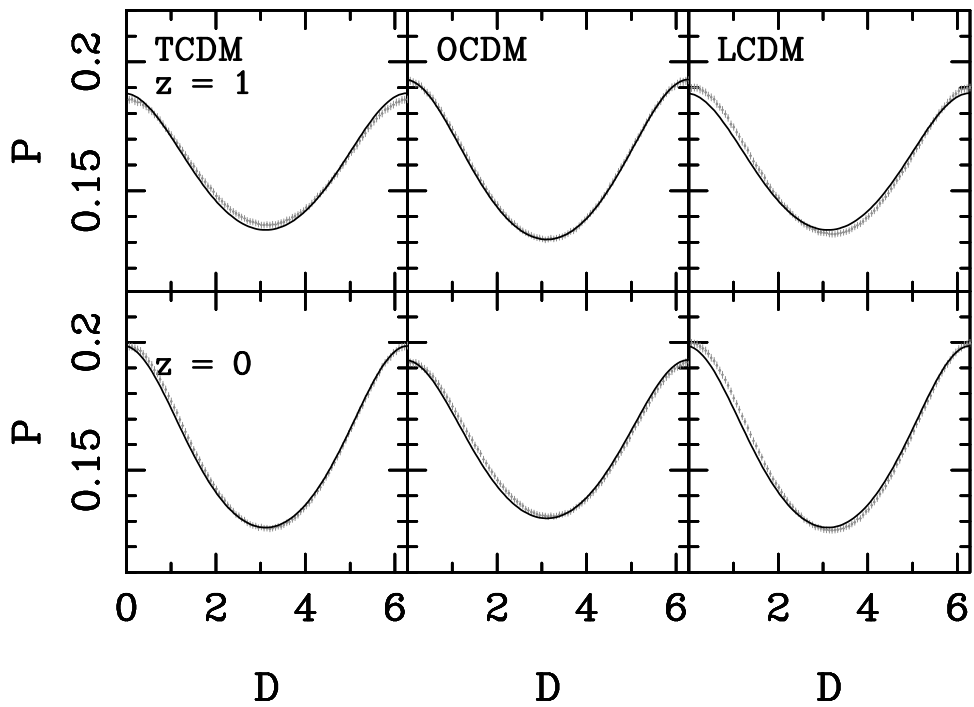}
\caption[]{Histograms $P(D)$ of phase differences $D$ are shown in grey  along with the
analytical model (\ref{eq:vm}) for three different CDM models
as functions of epoch.\label{cdm}}
\end{figure}

It is important to check whether deviations in the expansion rate or
non power-law initial power spectra would alter these conclusions.
Another ensemble of simulations used the power spectrum with CDM shape parameter
(Bardeen et al. 1986) $\Gamma = \Omega h = 0.225$.  
evolved in both an open ($\Omega_{m0} = 0.34$, OCDM), a flat
matter-dominated ($\Omega_{m0} = 1 $, TCDM), and a flat
cosmological-constant background ($\Omega_{m0} = 0.34$, $\Omega_{\lambda
0} = 0.66$, LCDM) All three ensembles were run to an amplitude
corresponding to $\sigma_8 = 0.93$.  A Hubble Constant $h = 2/3$ was used
in the simulation analysis.  These N-body runs had three realizations
each of $256^3$ particles in boxes of side 128 Mpc. The results are       
shown in figure \ref{cdm}; we found they are also nicely fit with the
von Mises distribution.
Results are shown also at $z$ = 1, which further varies the normalization
and background parameters, again with no significant deviation.

The excellent match of the distribution (\ref{eq:vm}) to the
results of detailed numerical simulations may appear surprising
given the very approximate nature of its derivation. But its
validity is reinforced by the fact that it is the maximum entropy
distribution on a circle for a fixed mean $\mu$ and fixed circular
dispersion \citep{mardia}. It should really therefore be regarded
as the circular equivalent of a Gaussian distribution, which has
maximal entropy for fixed mean and variance on the real line.

The universal behavior we have demonstrated will allow us in future to
discriminate between gravitationally-induced mode-coupling and other
forms, such as that induced by peculiar motions
\cite{melottetal98}. It should be pointed out, of course, that in a
realistic application to a survey of galaxies, further coupling
between the Fourier modes would arise due to the geometry and
selection function of the survey.  However, this need not be
problematic. If one knew the window function sufficiently
accurately its Fourier transform could be computed and a correction
applied to the complex Fourier components themselves.  \\

This work was supported by PPARC grant PPA/G/S/1999/00660; ALM
gratefully acknowledges the support of NSF through grant
AST-0070702, and computing support from the National 
Center for Supercomputing Applications.

\clearpage

\clearpage

\end{document}